\documentclass[english]{article}
\usepackage[T1]{fontenc}
\usepackage[latin9]{inputenc}
\usepackage{geometry}
\usepackage{array}
\usepackage{booktabs}
\usepackage{multirow}
\usepackage{amsmath}
\usepackage{amssymb}
\usepackage{graphicx}

\makeatletter

\providecommand{\tabularnewline}{\\}

\makeatother

\usepackage{babel}
\begin{document}
\title{Higher order mass aggregation terms in a nonlinear predator-prey model
maintain limit cycle stability in Saturn's F ring}
\author{Omar El Deeb \thanks{Department of Mathematics, University of Warwick, Coventry,
United Kingdom}}
\maketitle
\begin{abstract}
We consider a generic higher order mass aggregation
term for interactions between particles exhibiting oscillatory clumping
and disaggregation behavior in the F ring of Saturn, using a novel
predator-prey model that relates the mean mass aggregate (prey) and
the square of the relative dispersion velocity (predator) of the interacting
particles. The resulting cyclic dynamic behavior is demonstrated through
time series plots, phase portraits and their stroboscopic phase maps.

Employing an eigenvalue stability analysis of the Jacobian of the
system, we find out that there are two distinct regimes depending
on the exponent and the amplitude of the higher order interactions
of the nonlinear mass term. In particular, the system exhibits a limit
cycle oscillatory stable behavior for a range of values of these parameters
and a non-cyclic behavior for another range, separated by a curve
across which phase transitions would occur between the two regimes.
This shows that the observed clumping dynamics in Saturn's F ring,
corresponding to a limit cycle stability regime, can be systematically
maintained in presence of physical higher order mass aggregation terms
in the introduced model.
\end{abstract}

\section{Introduction}

Saturn, the sixth planet from the Sun in our solar system, is known
for its distinct appearance and its prominent ring system that spans
a diameter of approximately $280,000$ kilometers and sets it apart
from other planets. With its pale yellow hue, Saturn boasts a massive
atmosphere primarily composed of hydrogen and helium. Composed mainly
of ice particles and rocky debris, the rings consist of countless
individual ringlets, each with its own characteristics and unique
patterns. They consist of countless particles, ranging in size from
tiny grains to large chunks of ice and rock. The origin of Saturn's
rings remains a topic of scientific debate, but they are believed
to be the result of various processes, including the breakup of moons
or the remnants of a shattered moon that got too close to the planet
and was torn apart by tidal forces \cite{RingFormation1,RingFormation2}.
The rings are categorized into several main divisions, such as the
bright and broad A, B, and C rings, with the Cassini Division separating
the A and B rings. Additionally, fainter and narrower rings, such
as the F and G rings, exist within the system. They lie in the Roche
zone where the accretion is limited by Saturn\textquoteright s tidal
force and particle collisions \cite{Roche}. 

The F ring was initially discovered by the Pioneer 11 spacecraft in
1979 \cite{Pioneer11}. It is a narrow and faint ring located at about $3000$ km outside
the main ring system of Saturn, with a semi-major axis of about $140,000$
km. Analyses of the observed back-scattered signal from the Hubble
Space telescope, Voyager observations and other ground based observations
reveal that this ring is mainly composed of dust that constitutes
about $80-98\%$ of its particle fraction \cite{PsizeF}, in addition
to icy particles. These observations also reveal that it exhibits
a complex and ever-changing structure. It is exceptionally thin, with
an average width of about $30$ kilometers and consists of multiple
strands and braids of material that weave in and out, creating intricate
patterns and knots \cite{Voyager1,Voyager1.1}. These structures are
constantly evolving due to the gravitational interactions between
the ring particles and Saturn's moons \cite{Dynamics1}, particularly
the moon Prometheus, which orbits just inside the ring.

\begin{figure}
\centering{}\includegraphics[scale=0.25]{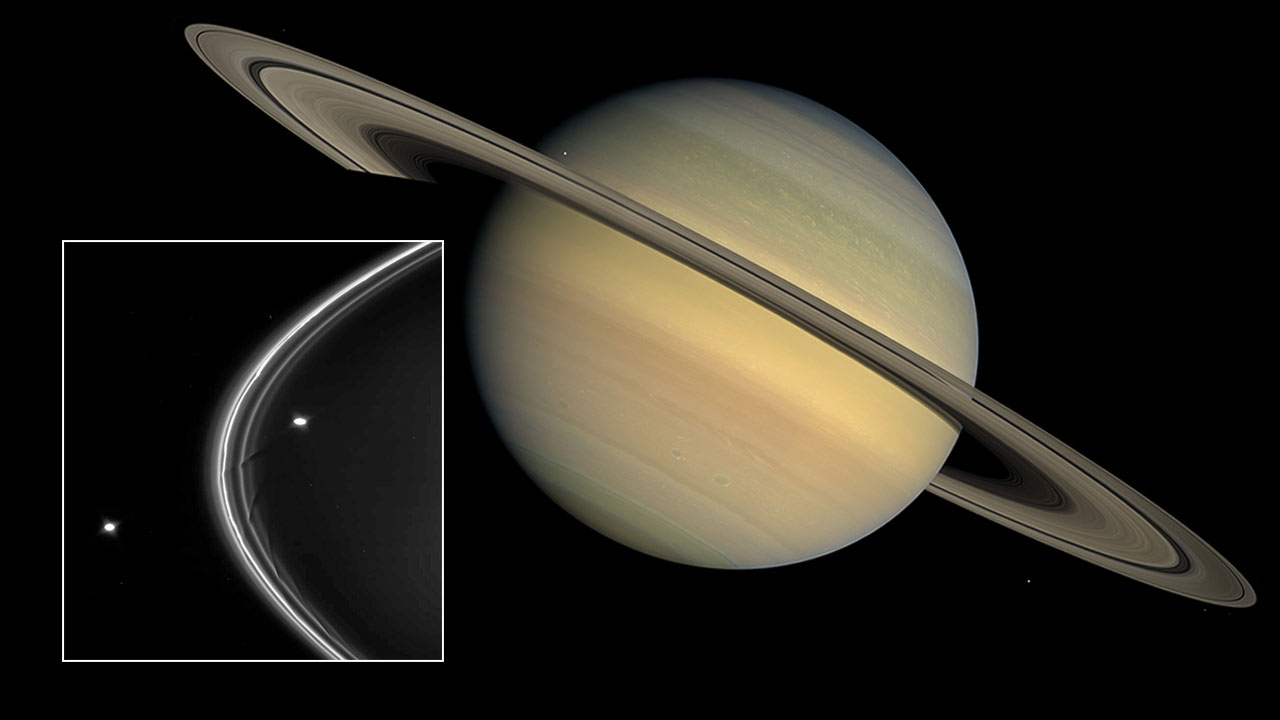}\caption{An image of Saturn and its rings. The zoomed image shows the
F ring with its nearest moons: Prometheus (inside ring) and Pandora (outside ring). Image credits: NASA/JPL/Space
Science Institute.}
\end{figure}

 The Cassini spacecraft had a primary
objective of studying Saturn, its rings and its moons in great detail.
Launched in 1997, Cassini arrived at Saturn in 2004 and embarked on
an extensive investigation of the planet and its surroundings.  
It studied Saturn's ring system, exploring its structure, origin,
and various divisions \cite{Cassini1,Cassini2} and also conducted
a thorough examination of Saturn's diverse moons, such as Titan and
Enceladus, investigating their geological features, compositions,
and interactions with Saturn's magnetosphere. 
Its images revealed multiple strands, braids, and clumps of material within the F ring,
showcasing its complex patterns. Related studies showed that the particles
had size distributions on several magnitudes across the ring, ranging
from micrometers up to the order of kilometer in size \cite{Psize1,Psize2}.
The spacecraft also captured the gravitational interactions between
the F ring and Saturn's shepherd moons, mainly the larger and nearer
Prometheus and then Pandora, which influenced its shape and confinement.
The two moons are shown in the zoomed image in Fig. (1) in the neighborhood
of the F ring \cite{SaturnImage}. Cassini detected temporary radial
structures called spokes composed of levitated dust particles and
observed denser regions known as knots within the F ring. Furthermore,
the mission unveiled the interactions between the F ring and nearby
moons, highlighting their impact on the ring's dynamics, leading to
disturbances, streamers, and localized concentrations of material
\cite{Clumps1,Clumps2}. Cassini's observations of heavily disturbed
areas during occultations reveal compelling indications of clustering
\cite{Occultations}. The observed clustering can be understood as a process of aggregation
followed by disaggregation, occurring over time spans ranging from
hours to weeks. The disruption of these clusters might be a result
of high-velocity collisions that cause fragmentation or the shedding
of loose fragments due to tidal forces. Additionally, the presence
of moons could contribute to the aggregation process by causing congestion
along their path, leading to increased collisions \cite{AggregFrag}.

Recent studies reported strong evidence of presence of moonlets within
the ring\textquoteright s bright core. They showed that ongoing gravitational
and collisional effects of small satellites combined with the perturbative
effects of Prometheus, the largest and closest moon in the neighborhood
of the F ring, are responsible for the resulting structure of the
F ring and its continual collisions and dynamics occurring on an almost
daily basis \cite{Dynamics}. The mean aggregate mass fluctuations
were explained through a system of coupled differential relating mass
aggregation and fragmentation with changes in relative dispersion
velocities of these particles through a predator-prey model \cite{PPmodel-Fring}. 

These types of models, also known as Lotka-Volterra models \cite{Lotka,volterra},
are mathematical models used to study the interactions between predator
and prey populations. The models consist of a set of differential
equations that describe the population dynamics of predators and prey
over time. They assume that the predator population's growth depends
on the availability of prey, while the prey population's growth is
influenced by predation. In a basic predator-prey model, there are
two variables: the population sizes of the predator species and the
prey species. Initially, an increase in prey population
leads to more resource availability for predators, causing the predator
population to grow. As the predator population increases, it exerts
more predation pressure on the prey, causing the prey population to
decline. With fewer prey available, the predator population eventually
decreases. As a result, the reduced predation allows the prey population
to recover, starting the cycle again. Predator-prey models were initially
introduced in the study of ecological systems, however, they are currently
employed in various models in fields that exhibit similar behavior,
ranging from particle dynamics to conservation biology, financial
markets and economics \cite{LV1,LV2,LV3,LV4,LV5,LV6,LV7,LV8}. 

However, predator-prey models introduced in the literature to study
the particle interaction dynamics in the rings of Saturn do not accommodate
for the possibility of higher order mass aggregation terms through
non-linear mass exponents in the model. Nonlinear
prey terms in the predator-prey model are important because they capture
the realistic dynamics of predator-prey interactions natural systems,
where the relationship between predators and prey dynamics is not
simply linear. Nonlinear prey terms can introduce various dynamics such as predator saturation,
oscillations, and stable coexistence of predator and prey populations.
By incorporating nonlinear prey terms, the predator-prey model can
better reflect the complexities of natural systems and provide insights
into the population dynamics and interactions in a more realistic
manner \cite{nonlin3}. Recent studies have shown
that the addition of nonlinear loss terms to the Lotka-Volterra model
in dust-forming plasmas stabilizes the oscillatory behavior in the
populations of particles in the plasma and better explains the observed
experimental behavior \cite{PlasmaNL}.

Our paper aims to broaden the scope of the previous predator-prey
model based simulations of Saturn's F ring and fill the current gap
in the literature by introducing a higher order interaction term for
mass aggregation with a generic nonlinear exponent, and study the
conditions under which the solutions would still exhibit limit cycle
oscillations around an equilibrium that can resemble the dynamics
corresponding to actual observations of clumping, spokes and knots across Saturn's F ring.

The paper is organized as follows: after the introduction section,
we construct our modified Lotka-Volterra model for the mean aggregate
mass dynamics in the F rings of Saturn in section 2. We explain the
methods required to analytically solve the system for some particular
cases, and the procedure to numerically solve the system for a generic
choice of parameters. In section 3, we present our results and discuss
them. We finally conclude in section 4.

\section{The Model}

The pioneering work in \cite{PPmodel-Fring} introduced a predator-prey
model for clumping triggered by moons in the rings of Saturn. 
The ring response periods vary on a time scale comparable
to the synodic period of the forcing of Prometheus. 
This model describes the long term behavior of the
ring dynamics based on gains and losses of the particle masses
and their relative velocities due to coagulation and fragmentation interplay.

\begin{center}
\begin{table}
\centering{}%
\begin{tabular}{>{\centering}p{1.6cm}>{\centering}p{2cm}>{\raggedright}m{8.8cm}}
\toprule 
\multirow{2}{1.6cm}{Parameter} & \multirow{2}{2cm}{\centering{}Value} & \multirow{2}{8.8cm}{\centering{}Comments}\tabularnewline\addlinespace[-0.1cm]
 &  & \tabularnewline
\midrule
\midrule 
$T_{orb}$ & \centering{}$1$ & \centering{}{\small{}All times are scaled with respect to $T_{orb}$}\tabularnewline\addlinespace[-0.1cm]
\midrule 
$\tau$ & \centering{}$0.1$ & \centering{}{\small{}Optical depth}\tabularnewline\addlinespace[-0.1cm]
\midrule 
$T$ & \centering{}$\frac{T_{orb}}{4\tau}$ & \centering{}{\small{}Estimated period of collisions}\tabularnewline
\midrule 
$\epsilon$ & \centering{}$0.6$ & \centering{}{\small{}Normal coefficient of restitution}\tabularnewline\addlinespace[-0.1cm]
\midrule 
$M_{0}$ & \centering{}$2\times10^{9}$g & \centering{}{\small{}Calculated for a $10$m aggregare with density
$\rho=0.5$ $g/cm^{3}$}\tabularnewline\addlinespace[-0.1cm]
\midrule 
$v_{esc}$ & \centering{}$0.5$ $m/s$ & \centering{}{\small{}The escape velolcity on the surface of $M_{0}$}\tabularnewline\addlinespace[-0.1cm]
\midrule 
$T_{syn}$ & $112$ $T_{orb}$ & \centering{}{\small{}The forcing period of Prometheus on the F ring}\tabularnewline\addlinespace[-0.1cm]
\midrule 
$A_{0}$ & $\in[0.1,1]$ & \centering{}{\small{}Amplitude of the moon driving force}\tabularnewline\addlinespace[-0.1cm]
\midrule 
$v_{th}$ & $v_{th}^{2}=\frac{A_{0}T_{syn}}{2\pi}$ & \centering{}{\small{}Threshold velocity for sticking}\tabularnewline\addlinespace[-0.1cm]
\bottomrule
\end{tabular}\caption{Parameters used for the numerical solutions of the modified predator-prey
model with non linear mass terms in the F ring.}
\end{table}
\par\end{center}

The coagulation term representing mass growth in \cite{PPmodel-Fring}
was approximated to grow in time in the form of $\frac{dM}{dt}=\frac{M}{T}$. 
In this approximation, $T=\frac{T_{orb}}{4\tau}$
is the collision period, $T_{orb}$ is the orbital period and $\tau$
is the optical depth at a specific location. In this paper, we consider
an additional modified generic growth term which includes a non-linear
mass dependence $kM^{n}$, with a nonlinear mass amplitude $k$ and
a degree $n$ to account for a more general and accurate approximation
and a possible higher order growth pattern in particle coagulation. The physical processes leading
to such nonlinearity could be attributed to numerous forms of interaction. Gravitational compression
 and inelastic collisions enhance local density and stabilize clumps, while gravitational instabilities
 cause small density perturbations to exponentially grow due to self-gravity. 
Shear reversal traps particles in areas of reduced relative velocity because of differential rotation,
 and swing amplification increases the amplitude of spiral density waves as they interact with the disk's rotation. 
Together, these mechanisms could introduce nonlinear terms resulting in the complex behavior seen in Saturn's rings.

Mass fragmentation occurs due to collisions of particles that reach
a relative velocity greater than the aggregates sticking threshold.
The mass loss due to fragmentation was found to be proportional to
the square of the relative velocity dispersion $v_{rel}$ and the
particle masses, such that $\frac{dM}{dt}=-\frac{V}{v_{th}^{2}}\frac{M}{T}$,
where $V=v_{rel}^{2}$ and $v_{th}$ is a parameter representing the
threshold velocity for sticking.

The temporal evolution of the relative velocity dispersion has a term
related to the losses caused by collisional dissipation. It is is
expressed by $\frac{dV}{dt}=-\frac{V(1-\epsilon^{2})}{T}$, where
$\epsilon$ is the normal coefficient of restitution. On the other
hand, local and non-local viscous stirrings contribute a velocity
growth term of the form $\frac{dV}{dt}=\frac{M^{2}}{M_{0}^{2}}\frac{v_{esc}^{2}}{T_{orb}}$
where $v_{esc}$ is the escape velocity from the surface of a reference
mass $M_{0}$. In addition, moon perturbations in the F ring, caused
by Prometheus, contribute an external forcing term with synodic angular
frequency $\omega=\frac{2\pi}{T_{syn}}$, and synodic period $T_{syn}$.

Consequently, taking into account all of these
contributions, we obtain a predator-prey model with a set of coupled nonlinear
differential equations that describe the time development of each
of $M(t)$ and $V(t)$.

\begin{equation}
\begin{cases}
\frac{dM}{dt}=\frac{M}{T}-\frac{V}{v_{th}^{2}}\frac{M}{T}+kM^{n}\\
\frac{dV}{dt}=-\frac{(1-\epsilon^{2})}{T}V+\frac{M^{2}}{M_{0}^{2}}\frac{v_{esc}^{2}}{T_{orb}}-A_{0}\cos(\omega t)
\end{cases}
\end{equation}

\begin{center}
\begin{figure}
\begin{centering}
\includegraphics[scale=0.75]{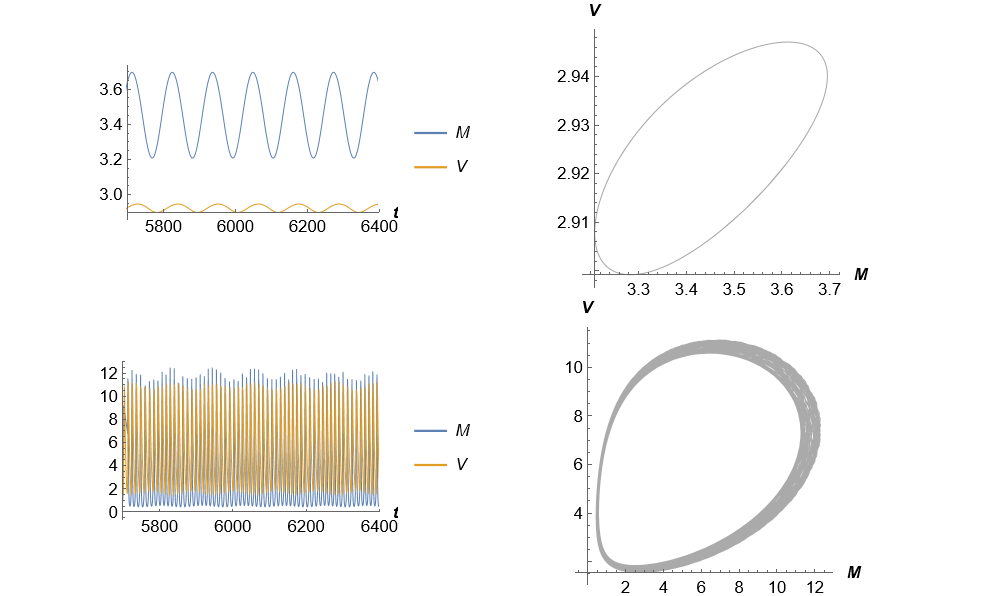}
\par\end{centering}
\caption{Time series plots and phase plots of $M$ and $V$, showing distinct dynamic behavior for $k=0.15, n=1.2$ (up) \& $k=0.6, n=1.3$ (bottom).  Parameters: $T_{syn}=112$ periods, $v_{esc}=0.5$ m/s, $M_{0}=2\times10^{9}g$, $A_{0}=\tau=0.1$, $\epsilon=0.6$ and initial conditions $M(0)=4.5\times10^{9}$g and $V(0)=3 m^{2}/s^{2}$.}
\end{figure}
\par\end{center}
 
The time in these equations is scaled with respect to the orbital
period, hence $T_{orb}\equiv1$. $A_{0}$ is the amplitude of the
moon forcing term, with $A_{0}=\frac{2\pi v_{th}^{2}}{T_{syn}}$.
A summary of all parameters and their numerical values is presented
in Table 1.

The system of coupled differential equations represents a predator-prey
model describing the mass fluctuations in the F ring of Saturn with
moon perturbations, with a generic non-linear higher order mass term.
In this context, the mean aggregate mass represents the prey population
whereas the square of the velocity dispersion represents the predator
population. The predator (relative velocity dispersion) grows larger
as the mass aggregates grow. But as it grows, more collisions would
occur on higher velocity scales causing increased fragmentation of
the aggregates, hence limiting its prey. This has similar qualitative
dynamics as ecological predator-prey systems, but with a unique quantitative
nature due to the simultaneous presence of a predator driving force,
a predator squared mass term and a generic prey non-linear mass term.

The equilibrium points of this system correspond to the solution of

\begin{equation}
\frac{dM}{dt}=\frac{dV}{dt}=0
\end{equation}

In the limiting case of a negligible driving force $A_{0}\rightarrow0$,
the equilibrium points, reached after amplitude death occurrence,
could be obtained from the solution of:

\begin{equation}
\begin{cases}
\frac{M}{T}\left(1-\frac{V}{v_{th}^{2}}+kTM^{n-1}\right)=0\\
-\frac{(1-\epsilon^{2})}{T}V+\frac{v_{esc}^{2}}{M_{0}^{2}}M^{2}=0
\end{cases}
\end{equation}

We can show that, upon reaching equilibrium, the square of the relative
velocity would be proportional to the mean aggregate mass:

\begin{equation}
V=\frac{Tv_{esc}^{2}}{(1-\epsilon^{2})M_{0}^{2}}M^{2}
\end{equation}

leading to the relation:

\begin{equation}
1-\frac{Tv_{esc}^{2}}{(1-\epsilon^{2})v_{th}^{2}M_{0}^{2}}M^{2}+kTM^{n-1}=0
\end{equation}

This is a polynomial of order $n-1$. It can be exactly solved for
some particular cases as in $n=1,2$ or $3$. The corresponding solutions
are given by:

\begin{equation}
\begin{cases}
M^{*}=\frac{(1+kT)(1-\epsilon^{2})v_{th}^{2}M_{0}^{2}}{Tv_{esc}^{2}}=\frac{v_{th}M_{0}}{v_{esc}}\sqrt{\left(k+\frac{1}{T}\right)\left(1-\epsilon^{2}\right)} & n=1\\
M^{*}=\frac{(1-\epsilon^{2})M_{0}^{2}\left(kT+\left(k^{2}T^{2}+\frac{4Tv_{esc}^{2}}{(1-\epsilon^{2})M_{0}^{2}v_{th}^{2}}\right)v_{th}^{2}\right)}{2Tv_{esc}^{2}} & n=2\\
M^{*}=\left(\frac{Tv_{esc}^{2}}{(1-\epsilon^{2})v_{th}^{2}M_{0}^{2}}-kT\right)^{-1/2} & n=3
\end{cases}
\end{equation}

Whereas the velocities could be simply obtained by substituting these
solution back in Eq. (4). However, analytical solutions are not attainable
in presence of the moon driving force, and for a generic $n$ case.

\begin{figure}
\begin{centering}
\includegraphics[scale=0.7]{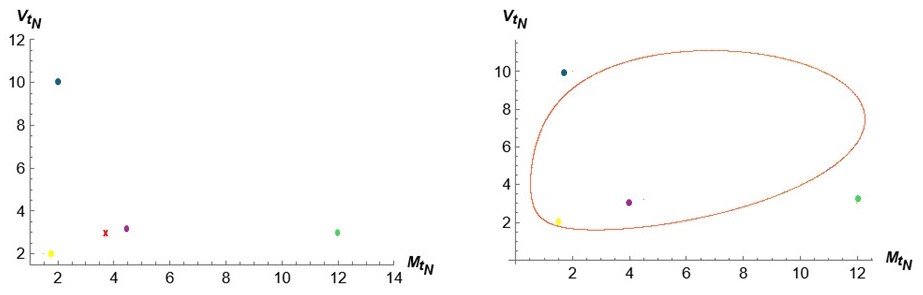}
\par\end{centering}
\caption{Stroboscopic phase space plots of $(M(t_{N}),V(t_{N}))$ at $t=N T_{syn}$  for $k=0.15, n=1.2$ (left) \& $k=0.6, n=1.3$ (right) for $4$ different initial conditions denoted by circular dots. On the left, all orbits reduce to a single point (red x), while on the right, all orbits fall into a closed cycle path.}
\end{figure}

To numerically solve this system, we adopt the following method. Stability
analysis requires linearising the modified predator-prey system with $A_{0}=0$ in
(3) around the equilibrium points. We determine the Jacobian of the
system represented by the matrix:

\begin{equation}
J=\begin{pmatrix}\frac{1}{T}\left(1-\frac{V^{*}}{v_{th}^{2}}\right)+nk\left(M^{*}\right)^{n-1} & -\frac{M^{*}}{v_{th}^{2}T}\\
\frac{2v_{esc}^{2}}{M_{0}^{2}}M^{*} & -\frac{1-\epsilon^{2}}{T}
\end{pmatrix}
\end{equation}

We determine the eigenvalues $\lambda$ of the Jacobian, which allow
us to determine the behavior and the stability of the system. For
a generic values of $n$ and $k$, we start by numerically solving
Eq. (5), to determine the corresponding values of $M$. We require
that $M>0$ for physical solutions. Then, for each $(n,k)$, we substitute
$n,$$k$ and $M$ in (7) to compute $\lambda$.

\section{Results and discussions}

We numerically solve the coupled system of non-linear differential
equations presented in Eq. (1) for various sets of parameters. We
plot the obtained solutions of the mean mass aggregate $M$ and the
square of the relative dispersion velocity $V$ as a function of time,
simultaneously in Fig. (2) for the following parameters: $T_{syn}=112$
orbital periods, $v_{esc}=0.5$ m/s, $M_{0}=2\times10^{9}g$, $A_{0}=0.1$,
$\epsilon=0.6$, $\tau=0.1$  and initial conditions $M(0)=4.5\times10^{9}$g and $V(0)=3$ $m^{2}/s^{2}$. This is done
in two distinct dynamic regimes, for $k=0.15$, $n=1.2$ (top) and for $k=0.6$, $n=1.3$ (bottom). We also plot their corresponding
phase space plot showing a continuous parametric plot of $V(t)$ with respect to $M(t)$.

\begin{figure}
\centering{}\includegraphics[scale=0.7]{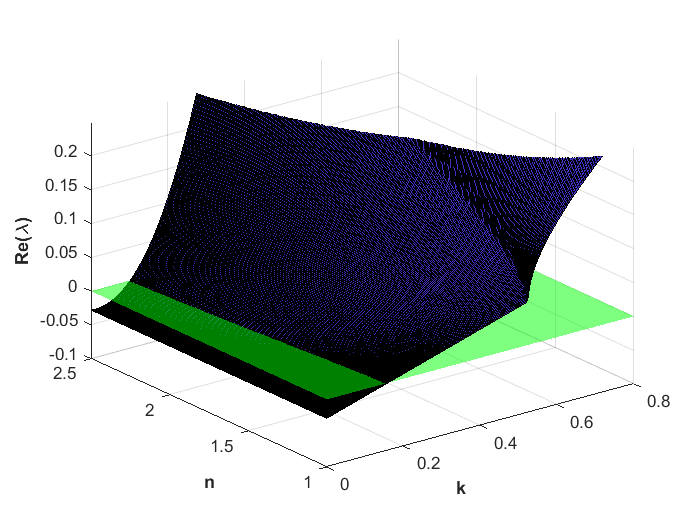}\caption{The real part of an eigenvalue of the Jacobian in Eq.7 ($A_{0}=0$), for parameters set to same values as  in Fig. 2, plotted in navy blue. The green plane corresponds to the $(n,k)$ plane or $Re(\lambda)=0$.}
\end{figure}

We observe that for a certain range of combinations of the nonlinear coupling $k$ and power $n$ , a periodic oscillation emerges
with a period identical to that of the perturbing synodic period. This is shown in Fig.2 (top). 
On the other hand, a more complicated dynamic behavior 
emerges for another range of values of $k$ and $n$. A sample plot is shown in Fig.2 (bottom). 
We can notice the presence of two patterns of 
frequencies, with rapid and slow oscillations.
The plots also show that the two variables under consideration are not in phase. 
The mean mass aggregate $M(t)$ (prey) peaks before the square of the dispersion
velocity $V(t)$ (predator), as expected for a predator lagging a
prey's population.

To understand the true nature of this dynamic behavior, we have to analyze the system 
by studying its stroboscopic map, with periodic snaphots of $M(t)$ and $V(t)$ taken at a 
$t_{N}=\frac{2\pi N}{\omega}=N T_{syn}$, for integer $N$. The corresponding phase space maps are plotted for $4$ initial conditions $(M(0),V(0)) = (1.5,2), (2,10), (4.5,3.2)$ and $(12,2)$, for $k=0.15, n=1.2$ (left) and $k=0.6, n=1.3$ (right). The figures show that for the former pair of nonlinear coupling and exponent, the orbit immediately 
reduces to a single point in the phase space after the first iteration, denoted by the red x in the figure. While for the latter parameters, using identical initial conditions, the corresponding stroboscopic phase plots reduce to a closed elliptical orbit, as shown in Fig. 3 (right).

With these stroboscopic maps, we can differentiate two types of periodic orbit in the full system. 
The case of $k=0.15, n=1.2$ shown in Fig. 3 (top) represents a periodic orbit which is Floquet-stable. 
It asymptotically tends toward the stable fixed point, representing a central periodic orbit with no limit cycle in this case.
Whereas for the case of $k=0.6, n=1.3$, we witness the emergence of a closed orbit limit cycle in the stroboscopic map. 
This represents the appearance of a limit torus in the full flow. 
The  phase transition fom a fixed point to a periodic orbit in the stoboscopic map indicates the occurrence of a quasi periodic Hopf bifurcation at a particular critical value of the pair $(k,n)$.

To further explore the critical values of $(k,n)$ that lead to Hopf bifurcation, 
we consider the unperturbed system (Eq. 3) with $A_{0}=0$ and analyze its eigenvalues.
Different $(k,n)$ parameterizations of the amplitude and exponent
parameters of the non-linear coagulation could lead to attracting stable fixed points or to limit cycle behviour. 
We classify the stability regions with respect to $(k,n)$ values  by employing a stability analysis of the Jacobian of the system.

After numerically solving Eq. (7), we present a three dimensional
plot of $Re(\lambda)$, the real part of the eigenvalue $\lambda$
of the Jacobian, with respect to $n$ and $k$ in Fig. (4). 
We obtain two distinct regions for $Re(\lambda)\neq0$. The $(n,k)$ plane corresponds to $Re(\lambda)=0$. It is plotted in green color in Fig. 4.
Solutions below the $(n,k)$ plane are due to attracting fixed points while those above the plane are due to limit cycle solutions. Transitions across
the line of intersection of $Re(\lambda)$ and the $(n,k)$ plane correspond to quasi periodic Hopf bifurcations, hence the birth of limit cycles, combined 
with the condition $Im(\lambda)\neq0$.

\begin{figure}
\centering{}\includegraphics[scale=0.66]{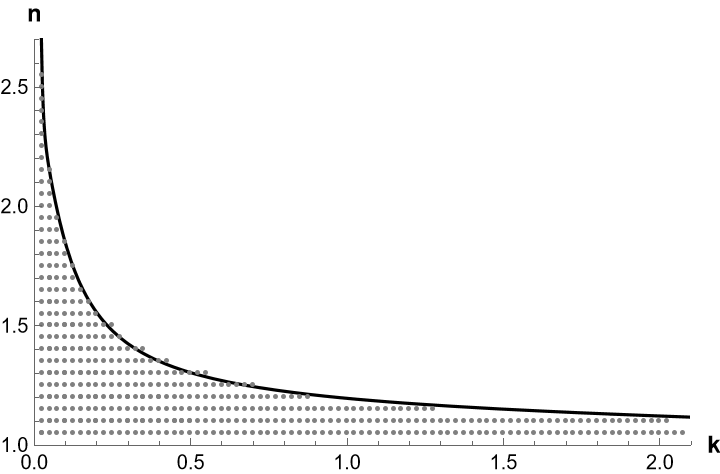}\caption{A scatter plot of $(k,n)$ values (gray dots) 
corresponding to $Im(\lambda)<0$ for $A_{0}=0$. The black line is a regression fit of the 
values at the edges of the scatter plot corresponding to $Im(\lambda)=0$. 
The region above the line (white region) corresponds to $Im(\lambda)>0$}
\end{figure}

To specify the solutions leading to $Im(\lambda)=0$, we choose
a large set of $(k,n$) such that $0\leq k\leq2.2$ and $1\leq n\leq2.6$
with an iteration of an increment of $0.025$ between consecutive
values of each parameter. We solve the modified Lotka-Volterra system
presented in Eq. (1) iteratively for all of these values, plot their
corresponding curves like those in Fig. (2), and graphically deduce
which ones correspond to $Im(\lambda)<0$. The
$(k,n)$ leading to this type of solutions are marked with gray dots
in the scatter plot in Fig. (5). The black line in that figure is
a plot of a polynomial regression fit of the edge values of the scatter
plot. It represents a boundary line emerging from $Im(\lambda)=0$.

Physically, our results imply that the actual
cyclic stability in Saturn's F rings can be maintained for a
range of couplings $k$ and exponents $n$ of a higher order mass
aggregation term. 

A sample image taken by Cassini \cite{Cassini} in Fig. (6)
reveals bright clumps of aggregated material on the edge of the F
ring in addition to fine structures and streams within the its core.
The process of disaggregation due to rapid collisions prohibits further growth of particles, and
keeps the system in cyclic equilibrium. This is inline with previous
studies \cite{Sizes} that analyzed particle size distribution
within the F ring and found out that they are characterized by an
inverse power law of the form:

\begin{equation}
n(s)=n(s_{0})\left(\frac{s}{s_{0}}\right)^{-q}
\end{equation}

where $s$ is the particle size, $s_{0}$ is a reference particle
size, and $n$ is the number of particles. 

Our results show that a nonlinear mass aggregation term can preserve all the dynamical features
corresponding to actual clumping, aggregations and disaggregations
that prohibit further mass growth in the ring and that the associated cyclic stability could be maintained for a well
defined region of values of the exponent and the amplitude of the
nonlinear mass growth term.

\begin{figure}
\centering{}\includegraphics[scale=0.4]{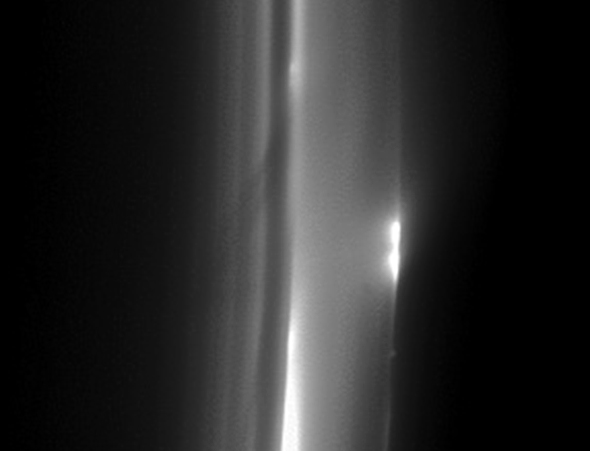}\caption{Actual bright clumps, fine structures and streams in Saturn's F rings.
Image Credit: NASA/JPL/SSI}
\end{figure}

\section{Conclusions}

The F ring is one of the most dynamic rings of Saturn, with variable
changes occurring on relatively short timescales. Prometheus has a
central role in triggering formations of streams, collisions and clumps.

In this study, a generic higher order mass aggregation term was considered
for particle interactions in the F ring of Saturn leading to aggregation,
disaggregation and clumping. A modified predator-prey model was used
to simulate the interactions of the prey and the predator variables,
namely: the mean mass aggregate and the relative dispersion velocities
of the particles. 

An eigenvalue stability analysis of the Jacobian of the system revealed
the existence of two distinct regimes depending on the exponent and
amplitude of the higher order interactions of the nonlinear mass term.
We explored the limit cycle oscillatory stable behavior for a range
of values of these parameters separated by a curve across which phase
transitions could occur into a distinct region of amplitude death
or instability. This work showed that the observed dynamical changes
in Saturn's F rings and their limit cycle stability features could
be systematically maintained in the presence of higher order mass
aggregation terms in the novel predator-prey model.

\section*{Acknowledgments}

The author acknowledges the valuable and insightful comments by two anonymous reviewers.


\begin{thebibliography}{10}
\bibitem{RingFormation1} R.M. Canup, Origin of Saturn\textquoteright s
rings and inner moons by mass removal from a lost Titan-sized satellite,
Nature 468, 943-946 (2010). doi: 10.1038/nature09661.

\bibitem{RingFormation2} K. Lumme, On the formation of Saturn's rings,
Astrophysics and Space Science 15, 404--414 (1972). doi: 10.1007/BF00649769.

\bibitem{Roche} M. Tiscareno, M. Hedman, J. Burns \& J. Castillo-Rogez,
Compositions and Origins of Outer Planet Systems: Insights from the
Roche Critical Density, The Astrophysical Journal Letters 765:2 (2013).
doi: 10.1088/2041-8205/765/2/L28.

\bibitem{Pioneer11} T. Gehrels, L.R. Baker, E. Beshore, et al., 
Imaging Photopolarimeter on Pioneer Saturn, Science 207 (4429), 434-439 (1980).
doi: 10.1126/science.207.4429.434.

\bibitem{PsizeF} F. Poulet, B. Sicardy, C. Dumas, et al., The crossings
of Saturn ring plane by the earth in 1995: ring thickness, Icarus
145, 147-165 (2000). doi: doi.org/10.1006/icar.1999.6314.

\bibitem{Voyager1} R. Hanel, B. Conrath, et al., Infrared Observations
of the Saturnian System from Voyager 1. Science 212, 4491, 192-200
(1981). doi: 10.1126/science.212.4491.1

\bibitem{Voyager1.1} B. Smith, L. Soderblom, et al., Encounter with
Saturn: Voyager 1 Imaging Science Results, Science, 212, 4491, 163-191
(1981). doi: 10.1126/science.212.4491.163.

\bibitem{Dynamics1} P. Goldreich, S. Tremaine, The Dynamics of Planetary
Rings, Annual Review of Astronomy and Astrophysics 20, 249-293 (1982).
doi: 10.1146/annurev.aa.20.090182.001341.


\bibitem{Cassini1} A. Nagy, A. Kliore, E. Marouf, et al., First results
from the ionospheric radio occultations of Saturn by the Cassini spacecraft,
Journal of Geophysical Research: Space Physics 111, A6 (2006). doi:
10.1029/2005JA011519.

\bibitem{Cassini2} D. Harland, Cassini at Saturn, Springer Praxis
Books, Praxis New York (2007). doi: 10.1007/978-0-387-73978-6.


\bibitem{Psize1} M. Hdeman, P. Nicholson, M. Showalter, et al., The
Christiansen Effect in Saturn\textquoteright s narrow dusty rings
and the spectral identification of clumps in the F ring, Icarus 215:2,
695-711 (2011). doi: 10.1016/j.icarus.2011.02.025.

\bibitem{Psize2} S. Vahidina, J. Cuzzi, M. Hedman, et al., Saturn\textquoteright s
F ring grains: Aggregates made of crystalline water ice, Icarus 215:2,
682-694 (2011). doi: 10.1016/j.icarus.2011.04.011.

\bibitem{SaturnImage} Science, Satellite smashup created Saturn's
narrow F ring (2015), Accessed on: June 12, 2023. Available at: https://www.science.org/content/article/satellite-smashup-created-saturns-narrow-f-ring.

\bibitem{Clumps1} L. Esposito, B. Meinke, J. Colwell, et al., Moonlets
and clumps in Saturn's F ring, Icarus 194:1, 278-289 (2008). doi:
10.1016/j.icarus.2007.10.001.

\bibitem{Clumps2} N. Albers, M. Sremcevic, J. Colwell \& L. Esposito,
Saturn\textquoteright s F ring as seen by Cassini UVIS: Kinematics
and statistics, Icarus 217:1, 367-388 (2012). doi: 10.1016/j.icarus.2011.11.016.

\bibitem{Occultations} B. Meinke, L. Esposito, N. Albers \& M. Sremcevic,
Classification of F ring features observed in Cassini UVIS occultations
218:1, 545-554 (2012). doi: 10.1016/j.icarus.2011.12.020.

\bibitem{AggregFrag} N. Brilliantov, P.L. Krapivsky, A. Bodrova \&
J. Schmidt, Size distribution of particles in Saturn\textquoteright s
rings from aggregation and fragmentation, Proceedings of the National
Academy of Science 112:31, 9536-9541 (2015). doi: 10.1073/pnas.1503957112.

\bibitem{Dynamics} C. Murray, K. Beurle, N. Cooper, et al., The determination
of the structure of Saturn\textquoteright s F ring by nearby moonlets.
Nature 453, 739--744 (2008). doi: 10.1038/nature06999.

\bibitem{PPmodel-Fring} L. Esposito, N. Albers, B. Meinke, et al.,
A predator--prey model for moon-triggered clumping in Saturn\textquoteright s
rings, Icarus 217:1, 103-114. doi: 10.1016/j.icarus.2011.09.029.

\bibitem{Lotka} A. Lotka, Analytical Note on Certain Rhythmic Relations
in Organic Systems, Proceedings of the National Academy of Science
6:7, 410-415 (1920). doi: 10.1073/pnas.6.7.4.10.

\bibitem{volterra} V. Volterra, Variations and fluctuations of the
number of individuals in animal species living together, Animal Ecology,
McGraw-Hill 412-433 (1931).

\bibitem{LV1} P. Wangersky, Lotka-Volterra Population Models, Annual
Review of Ecology and Systematics 9, 189-218 (1978). doi:10.1146/annurev.es.09.110178.001201.

\bibitem{LV2} L. Frachebourg, P.L. Krapivsky \& E. Ben-Naim, Spatial
organization in cyclic Lotka-Volterra systems, Physical Review E 54:6,
6186 (1996). doi:10.1103/PhysRevE.54.6186.

\bibitem{LV3} G. Dibeh \& O. El Deeb, Synchronization in a market model with time delays, 
arXiv:2405.00046 (2024). doi:10.48550/arXiv.2405.00046.

\bibitem{LV4} A. Chakrabarti, Stochastic Lotka--Volterra equations:
A model of lagged diffusion of technology in an interconnected world,
Physica A: Statistical Mechanics and Its Applications 442, 214-223
(2016). doi:10.1016/j.physa.2015.09.030.

\bibitem{LV5} Y. Marinkas, R. White \& S. Walsh, Lotka--Volterra
signals in ASEAN currency exchange rates, Physica A: Statistical Mechanics
and Its Applications 545, 123743 (2020). doi:10.1016/j.physa.2019.123743.

\bibitem{LV6} G. Dibeh, Contagion effects in a chartist--fundamentalist
model with time delays, Physica A: Statistical Mechanics and Its Applications
382:1, 52-57 (2007). doi:10.1016/j.physa.2007.02.007.

\bibitem{LV7} T. Ray, L. Moseley \& N. Jan, A Predator--Prey Model
with Genetics: Transition to a Self-Organized Critical State, International
Journal of Modern Physics C 9:5, 701-710 (1998), doi: 10.1142/S0129183198000601.

\bibitem{LV8} W. Just, E. Reibold, H. Benner et al., Limits of time-delayed
feedback control, Physics Letters A 254:3-4, 158-164 (1999). doi:
10.1016/S0375-9601(99)00113-9.

\bibitem{nonlin3} K. Owolabi, Computational dynamics of predator-prey
model with the power-law kernel, Results in Physics 21, 103810 (2021).
doi: 10.1016/j.rinp.2020.103810.

\bibitem{PlasmaNL} A.E. Ross \& D.R. McKenzie, Predator-prey dynamics
stabilized by nonlinearity explain oscillations in dust-forming plasmas,
Scientific Reports 6, 24040 (2016). doi: 10.1038/srep24040.

\bibitem{NonLin1} Q. Shu \& J. Xie, Stability and bifurcation analysis
of discrete predator--prey model with nonlinear prey harvesting and
prey refuge, Mathematical Methods in the Applied Sciences 45:7 (2021),
3589-3604. doi:10.1002/mma.8005.

\bibitem{NonLin2} Y. Tao, Global existence of classical solutions
to a predator--prey model with nonlinear prey-taxis, Nonlinear Analysis:
Real World Applications 11 (2010), 2056-2064. doi:10.1016/j.nonrwa.2009.05.005.

\bibitem{Bifurcation} X. Liu, T. Zhang, X. Meng et al., Turing--Hopf
bifurcations in a predator--prey model with herd behavior, quadratic
mortality and prey-taxis, Physica A: Statistical Mechanics and its
Applications 496, 446-460 (2018). doi: 10.1016/j.physa.2018.01.006.


\bibitem{Cassini} J. Major, Saturn's fluctuating F ring, Phys.org,
(2012). Availble at: https://phys.org/news/2012-11-saturn-fluctuating.html.
Accessed on: June 23, 2023.

\bibitem{Sizes} C. Murray \& R. French, The F Ring of Saturn. In:
M. Tiscareno \& C. Murray, Planetary Ring Systems: Properties, Structure,
and Evolution, Cambridge: Cambridge Univ. Press, 338-362 (2018). doi:
10.1017/9781316286791.013.



\end{thebibliography}
\end{document}